\documentclass[fleqn]{jaes}

\usepackage{newtxtext,newtxmath}
\usepackage{bm}
\usepackage[hyphens]{url}
 \usepackage[verbose]{placeins}
 
\newcommand{\myfW}{\ensuremath{{f_{\boldsymbol{W}}}}}
\newcommand{\myg}{\ensuremath{\boldsymbol{g}}}
\newcommand{\myPhi}{\ensuremath{\boldsymbol{\Phi}}}
\newcommand{\mytheta}{\ensuremath{\boldsymbol{\theta}}}
\newcommand{\mythetatilde}{\ensuremath{\boldsymbol{\widetilde{\theta}}}}
\newcommand{\myW}{\ensuremath{\mathbf{W}}}
\newcommand{\myx}{\ensuremath{\boldsymbol{x}}}
\newcommand{\myy}{\ensuremath{\boldsymbol{y}}}
\newcommand{\mypsi}{\ensuremath{\boldsymbol{\psi}}}

\begin{document}


\title{Mesostructures: Beyond Spectrogram Loss \\
in Differentiable Time--Frequency Analysis}

\authorgroup{
\author{CYRUS VAHIDI\textsuperscript{1},}
\author{HAN HAN\textsuperscript{2},}
\author{CHANGHONG WANG\textsuperscript{2},}\\
\author{MATHIEU LAGRANGE\textsuperscript{2},}
\author{GYÖRGY FAZEKAS\textsuperscript{1}},
AND \author{VINCENT LOSTANLEN\textsuperscript{2}}
\email{(c.vahidi@qmul.ac.uk)}
\affil{\textsuperscript{1}Centre for Digital Music, Queen Mary University of London, United Kingdom \\
\textsuperscript{2}Nantes Universit\'e,  \'Ecole Centrale Nantes, CNRS, LS2N, UMR 6004, F-44000 Nantes, France}
}

\abstract{%
Computer musicians refer to mesostructures as the intermediate levels of articulation between the microstructure of waveshapes and the macrostructure of musical forms. Examples of mesostructures include melody, arpeggios, syncopation, polyphonic grouping, and textural contrast.
Despite their central role in musical expression, they have received limited attention in deep learning.
Currently, autoencoders and neural audio synthesizers are only trained and evaluated at the scale of microstructure: i.e., local amplitude variations up to 100 milliseconds or so.
In this paper, we formulate and address the problem of mesostructural audio modeling via a composition of a differentiable arpeggiator and time-frequency scattering. We empirically demonstrate that time--frequency scattering serves as a differentiable model of similarity between synthesis parameters that govern mesostructure.
By exposing the sensitivity of short-time spectral distances to time alignment, we motivate the need for a time-invariant and multiscale differentiable time--frequency model of similarity at the level of both local spectra and spectrotemporal modulations. 
}

\maketitle

\section{INTRODUCTION}
\label{sec:intro}

\subsection{Differentiable time--frequency analysis}
Time--frequency representations (TFR) such as the short-term Fourier transform (STFT) or constant-Q transform (CQT) play a key role in music signal processing \cite{schorkhuber2010constant, muller2015fundamentals} as they can demodulate the phase of slowly varying complex tones.
As a consequence, any two sounds \myx{} and \myy{} with equal TFR magnitudes (i.e., spectrograms) are heard as the same by human listeners, even though the underlying waveforms may differ.
For this reason, spectrograms can not only serve for visualization, but also for similarity retrieval.
Denoting the spectrogram operator by \myPhi{}, the Euclidean distance $\Vert \myPhi(\myy) - \myPhi(\myx)\Vert_2$ is much more informative than the waveform distance $\Vert \myy - \myx\Vert_2$, since the waveform distance diverges quickly even when phase differences are small.

In recent years, existing algorithms for STFT and CQT have been ported to deep learning frameworks such as PyTorch, TensorFlow, MXNet, and JAX \cite{cheuk2020nnaudio, andreux2018music, yang2021torchaudio}. 
By doing so, the developers have taken advantage of the paradigm of differentiable programming, defined as the ability to compute the gradient of mathematical functions by means of reverse-mode automatic differentiation.
In the context of audio processing, differentiable programming may serve to train a neural network for audio encoding, decoding, or both.
Hence, we may coin the umbrella term \emph{differentiable time--frequency analysis} (DTFA) to describe an emerging subfield of deep learning in which stochastic gradient descent involves a composition of neural network layers as well as TFR. 
Previously, TFR were largely restricted to analysis frontends, but now play an integral part in learning architectures for audio generation.

The simplest example of DTFA is autoencoding.
Given an input waveform \myx{}, the autoencoder is a neural network architecture $f$ with weights \myW{}, which returns another waveform \myy{} \cite{engel2019ddsp, zeghidour2021soundstream}.
During training, the neural network \myfW{} aims to minimize the following loss function:
\begin{equation}
\mathcal{L}_{\myx}(\myW) = \Vert (\myPhi \circ \myfW)(\myx) - \myPhi(\myx) \Vert_2,
\end{equation}
on average over every sample \myx{} in an unlabeled dataset.
The function above is known as \emph{spectrogram loss} because \myPhi{} maps \myx{} and \myy{} to the time--frequency domain.

Another example of DTFA is found in audio restoration.
This time, the input of \myfW{} is not \myx{} itself but some degraded version $h(\myx)$ --- noisy or bandlimited, for example \cite{manocha2020differentiable, su2021bandwidth}.
The goal of \myfW{} is to invert the degradation operator $\boldsymbol{h}$ by producing a restored sound $(\myfW \circ \boldsymbol{h})(\myx)$ which is close to \myx{} in terms of spectrogram loss:
\begin{equation}
\mathcal{L}_{\myx}(\myW) = \Vert (\myPhi \circ \myfW \circ h)(\myx) - \myPhi(\myx) \Vert_2.
\end{equation}

Thirdly, DTFA may serve for sound matching, also known as synthesizer parameter inversion \cite{engel2019ddsp, esling2019flow, masuda2021synthesizer}.
Given a parametric synthesizer \myg{} and an audio query \myx{}, this task consists in retrieving the parameter setting $\mytheta$ such that $\myy=\myg(\mytheta)$ resembles \myx{}.
In practice, sound matching may be trained on synthetic data by sampling $\mytheta$ at random, generating $\myx=\myg(\mytheta)$, and measuring the spectrogram loss between \myx{} and \myy{}:
\begin{equation}
\mathcal{L}_{\mytheta}(\myW) = \Vert (\myPhi \circ \myg \circ \myfW \circ \myg)(\mytheta) - (\myPhi \circ \myg)(\mytheta) \Vert_2.
\end{equation}

\subsection{Shortcomings of spectrogram loss}
Despite its proven merits for generative audio modeling, spectrogram loss suffers from counterintuitive properties when events are unaligned in time or pitch \cite{turian2020m}.
Although a low spectrogram distance implies a judgment of high perceptual similarity, the converse is not true: one can find examples in which $\myPhi(\myx)$ is far from $\myPhi(\myy)$ yet judged musically similar by a human listener.
First, $\myPhi$ is only sensitive to time shifts up to the scale $T$ of the spectrogram window; i.e., around 10--100 milliseconds.
In the case of autoencoding, if $\myfW(\myx)(t) = \myx(t - \tau)$ with $\tau \gg T$, $\mathcal{L}_{\myx}(\myW)$ may be as large as $2 \Vert \myPhi(\myx)\Vert_2$ even though the output of $\myfW{}$ would be easily realigned onto $\myx{}$ by cross-correlation.
In the case of audio restoration of pitched sounds, listeners are more sensitive to artifacts near the onset (e.g., pre-echo) \cite{brandenburg1999mp3}, even though most of the spectrogram energy is contained in the sustain and release parts of the temporal profile.

Lastly, in the case of sound matching, certain synthesizers contain parameters which 
govern periodic structures at larger time scales while being independent of local spectral variations.
In additive synthesis, periodic modulation techniques such as vibrato, tremolo, or trill have a ``rate’’ parameter which is neither predictable from isolated spectrogram frames, nor reducible to a sequence of discrete sound events. 
A small perturbation to synthesis parameters  of $\varepsilon$ will induce a $\myg{}(\mytheta{}+\varepsilon)$ globally dilated or compressed but locally misaligned  in time, rendering $\Vert (\myPhi{} \circ \myg){(\mytheta{}+ \varepsilon)} - (\myPhi{} \circ \myg){(\mytheta{})} \Vert$ not indicative of the magnitude of $\varepsilon$. Modular synthesizers shape sound via an interaction between control modules (sequencers, function generator) and sound processing \& generating modules (oscillators, filters, waveshapers) \cite{subotnick1970use}.
In a ``patch``, sequencers determine the playback speed and actuate events, while amplitude envelopes, oscillator waveshapes and filters sculpt the timbre.
Changing the clock speed of a patch would cause events to be unaligned in time, but not alter the spectral composition of isolated events. 
Therefore comparison of timbre similarity is no longer possible at the time scale of isolated spectrogram frames.


\subsection{Musical timescales: micro, meso, macro}
The shortcomings of modelling music similarity solely at the microscale of short-term spectra is exemplified by the terminology of musical structure used in algorithmic composition.
Computer musicians refer to musical structures at a hierarchy of time scales. At one end is the \emph{micro scale}; from sound particles of few samples up to the milliseconds of short-term spectral analysis \cite{roads2004microsound}. 
Further up the hierarchy of time is the \emph{meso scale}; structures from that emerge from the grouping of sound objects and their complex spectrotemporal evolution \cite{roads2014rhythmic}. While the \emph{macro scale} broadly includes the arrangement of a whole composition or performance.  
Curtis Roads outlines the challenge of coherently modeling multiscale structures in algorithmic composition \cite{roads2012grains}. 
In granular synthesis, microstructure arises from individual grains, while their rate of playback forms texture clouds at the level of mesostructure.
Beyond the micro scale and spectrogram analysis are sound structures that emerge from complex spectral and temporal envelopes, such as sound textures and instrumental playing techniques \cite{lostanlen2019fourier}.

\subsection{Contributions}
In this paper, we pave the way towards differentiable time--frequency analysis of mesostructure.
The key idea is to compute a 2D wavelet decomposition (``scattering'') in the time--frequency domain for a sound $\myx$.
The result, named joint time--frequency scattering transform (JTFS), is sensitive to relative time lags and frequency intervals between musical events.
Meanwhile, JTFS remains stable to global time shifts: going back to the example of autoencoding, $\myfW(\myx)(t) = \myx(t - \tau)$ leads to $(\myPhi \circ \myfW)(\myx) \approx \myPhi(\myx)$, in line with human perception.

To illustrate the potential of JTFS in DTFA, we present an example of differentiable sound matching in which microscale distance is a poor indicator of parameter distance.
In our example, the target sound $\myx = \myg(\mytheta)$ is an arpeggio of short glissandi events (``chirplets'') which spans a scale of two octaves.
The two unknowns of the problem are the number of chirplets per unit of time and the total duration of the arpeggio.
We show that it is possible to retrieve these two unknowns without any feature engineering, simply by formulating a least squares inverse problem in JTFS space of the form:
\begin{align}
\mytheta^{*} =&
\ \mathrm{arg}\,\min_{\mythetatilde} \mathcal{L}_{\mytheta}(\mythetatilde) \nonumber \\=& \ \mathrm{arg}\,\min_{\mythetatilde} \Vert (\myPhi \circ \myg)(\mythetatilde) - (\myPhi \circ \myg)(\mytheta) \Vert_2^2
\end{align}

Intuitively, for the inverse problem above to be solvable by gradient descent, the gradient of $\mathcal{L}_{\mytheta}$ should point towards $\mytheta$ when evaluated at any initial guess $\mythetatilde$.
Our main finding is that such is the case if $\myPhi$ is JTFS, but not if $\myPhi$ is the multi-scale spectrogram (MSS).
Moreover, we find that the gradient of $\mathcal{L}_{\mytheta}$ remains informative even if the target sound is subject to random time lags of several hundred milliseconds.
To explain this discrepancy, we define the concept of \emph{differentiable mesostructural operator} as yielding the Jacobian matrix of $(\myPhi \circ \myx)$ at $\mythetatilde$, i.e., the composition between audio synthesis and JTFS analysis at the parameter setting of interest.
This concept is not limited to sound matching but also finds equivalents when training neural networks for autoencoding and audio restoration.

\begin{figure*}[t!]
\includegraphics[width=\linewidth]{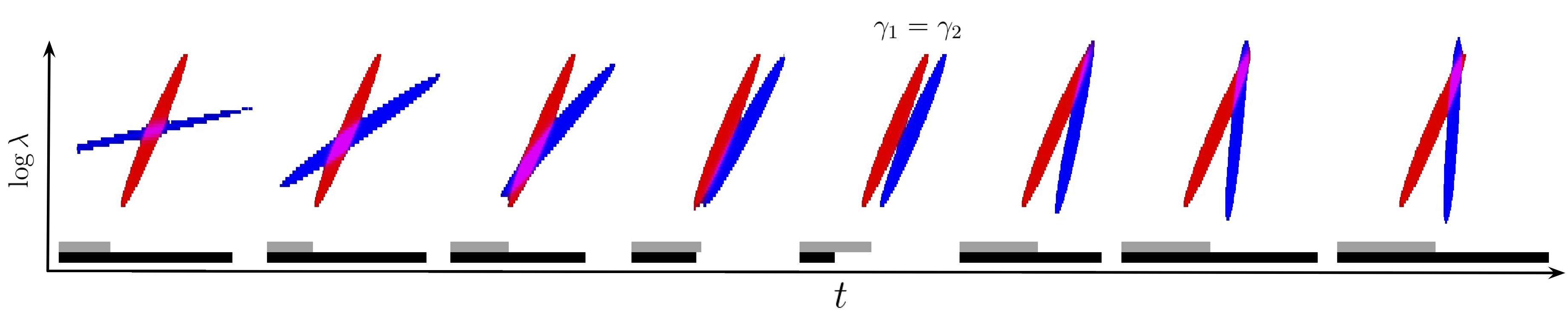}
\caption{Illustration of chirps overlapping in time and log--frequency. 
The red chirps are of equal chirp rate $\gamma$. 
The blue chirps are displaced in time from red and of increasing $\gamma$ (left to right). 
The bars indicate the distance between two chirps in the multiscale spectrogram (grey) and time--frequency scattering (black) domains, respectively.
We observe that when the chirp rates $\gamma$ governing \emph{mesostructure} are equal, the JTFS distance is at a minimum, while spectrogram distance is around its maximum. JTFS distance correlates well with distance in $\gamma$. 
We give a more detailed discussion of the importance of a time-invariant differentiable mesostructural operator in Section \ref{sec:dmo}.}
\label{fig:chirps}
\end{figure*}

We release a differentiable implementation of JTFS in Kymatio v0.4\footnote{Kymatio v0.4: \url{https://github.com/kymatio/kymatio}}, an open-source software for DTFA on GPU which is interoperable with modern deep learning libraries \cite{andreux2020kymatio}. 
To encourage reproducibility of numerical experiments, we supplement this paper with open-source code\footnote{Experiments repository: \url{ https://github.com/cyrusvahidi/meso-dtfa}}.


\section{MOTIVATING EXAMPLE} \label{sec:am_chirp}

\subsection{Comparing time-delayed chirps}
Fig. (\ref{fig:chirps}) illustrates the challenge in DTFA of reliably computing similarity between chirps synthesized by \myg. 
In the example, the first-order moments of two chirps in the time--frequency domain are equal, regardless of FM rate. 
Consider two chirps that are displaced from one another in time. 
Their spectrogram distance is at a maximum when the mesostructure is identical, i.e. the FM rates are equal and the two signals are disjoint. 
As the FM rate increases, the two chirps overlap in the time--frequency domain, resulting in a reduction of the spectrogram distance that does not correlate with correct prediction of \mytheta{}. 
The spectrogram loss changes little as $\gamma$ is varied. 
Moreover, local micro segments of a chirp are periodically shifted in \emph{both} time and frequency under $\gamma$, implying that comparison of microstructure is an inadequate indicator of similarity.
A possible solution would be to dynamically realign the chirps, however this operation is numerically unstable and not differentiable.
In the following sections, we outline a differentiable operator that is capable of modelling distance in \mytheta{} and stable to time shifts. A representation that is well-equipped to disentangle these three factors of variability should provide neighbourhood distance metrics in acoustic space that reflect distance in parameter space. 

\subsection{Chirplet synthesizer}
A chirplet is a short sound event which produces a diagonal line in the time--frequency plane.
Generally speaking, chirplets follow an equation of the form $\myx{}(t) = \boldsymbol{a}(t) \cos(2\pi \boldsymbol{\varphi}(t))$ where $\boldsymbol{a}$ and $\boldsymbol{\varphi}$ denote instantaneous amplitude and phase respectively.
In this paper, we generate chirplets whose instantaneous frequency grows exponentially with time, so that their perceived pitch (roughly proportional to log-frequency) grows linearly.
We parametrize this frequency modulation (FM) in terms of a chirp rate $\gamma$, measured in octaves per second.
Denoting by $f_\mathrm{c}$ the instantaneous frequency of the chirplet at its onset, we obtain:
\begin{equation}
\boldsymbol{\varphi}(t) = \dfrac{f_\mathrm{c}}{\gamma \log 2} 2^{\gamma t}.
\end{equation}
Then, we define the instantaneous amplitude $\boldsymbol{a}$ of the chirplet as the half-period of a sine function, over a time support of $\delta^{\mathrm{t}}$.
We parameterise this half-period in terms of amplitude modulation (AM) frequency $f_\mathrm{m}=\frac{1}{2}\delta^{\mathrm{t}}$.
Hence:
\begin{equation}
\boldsymbol{a}(t) = \sin(2\pi f_{\mathrm{m}} t)
\textrm{ if }0\leq f_{\mathrm{m}}t<\tfrac{1}{2}\textrm{ and }0\textrm{ otherwise.}
\end{equation}
At its offset, the instantaneous frequency of the chirplet is equal to $f_\mathrm{m} = f_\mathrm{c} 2^{\gamma \delta^{\mathrm{t}}} = f_\mathrm{m} 2^{\gamma/f_\mathrm{m}}$.
We use the notation $\mytheta$ as a shorthand for the AM/FM tuple $(f_{\mathrm{m}}, \gamma)$.

\subsection{Differentiable arpeggiator}
We now define an ascending ``arpeggio'' such that the offset of the previous event coincides with the onset of the next event in the time--frequency domain.
To do so, we shift the chirplet by $n \delta^{t}$ in time and multiply its phase by $2^{n \delta^{\mathrm{f}}} = 2^{n \gamma \delta^{\mathrm{t}}}$ for integer $n$.
Lastly, we apply a global temporal envelope to the arpeggio, by means of a Gaussian window $(t\mapsto\boldsymbol{\phi}_{w}(\gamma t)/\gamma)$ of width $\gamma w$ where the bandwidth parameter $w$ is expressed in octaves.
Hence:
\begin{align}
\boldsymbol{x}(t) &=
\dfrac{1}{\gamma}\boldsymbol{\phi}_{w}(\gamma t)
\sum_{n=-\infty}^{+\infty}
\boldsymbol{a}\left(t-\frac{n}{f_\mathrm{m}}\right)
\cos\left(
2^{\gamma\frac{n}{f_{\mathrm{m}}}}
\boldsymbol{\varphi}\left(
t - \dfrac{n}{f_{\mathrm{m}}
}
\right)
\right)
\nonumber \\
&= \myg_{\mytheta}(t)\textrm{, where }\mytheta=(f_{\mathrm{m}}, \gamma).
\end{align}
In the equation above, the number of events with non-negligible energy is proportional to:
\begin{equation}
\nu(\mytheta) = \dfrac{f_{\mathrm{m}}w}{\gamma},
\end{equation}
which is not necessarily an integer number since it varies continuously with respect to $\mytheta$.
Here we see that our parametric model $\boldsymbol{g}$, despite being very simple, controls an auditory sensation whose definition only makes sense at the mesoscale: namely, the number of notes $\nu$ in the arpeggio that form a sequential stream.
Furthermore, this number results from the entanglement between AM ($f_{\mathrm{m}}$) and FM ($\gamma$) and would remain unchanged after time shifts (replacing $t$ by $(t-\tau)$) or frequency transposition (varying $f_\mathrm{c}$).
Thus, although the differentiable arpeggiator has limited flexibility, we believe that is offers an insightful test bed for the DTFA of mesostructure.

\section{TIME--FREQUENCY SCATTERING} \label{sec:jtfs}
Joint time--frequency scattering (JTFS) is a convolutional operator in the time--frequency domain \cite{Anden2019joint}. 
Via two-dimensional wavelet filters applied in the time--frequency domain at various scales and rates, JTFS extracts multiscale spectrotemporal modulations from digital audio. 
When used as a frontend to a 2D convolutional neural network, JTFS enables state-of-the-art musical instrument classification with limited annotated training data \cite{muradeli2022differentiable}. 
Florian Hecker's compositions, e.g \emph{FAVN} in 2016, mark JTFS's capability of computer music resynthesis (see a full list of compositions from \cite{lostanlenflorian19}).

\begin{figure}
    \centering
    \includegraphics[width=\columnwidth]{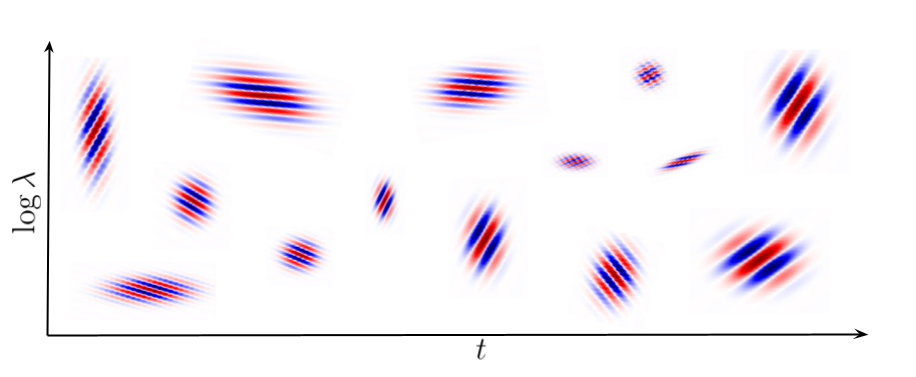}
    \caption{Illustration of the shape of 2D time--frequency wavelets (second-order JTFS). 
    Red and blue indicate higher positive and lower negative values (resp.). 
    Each pattern shows the response of the real part of two-dimensional filters that arise from the outer product between 1D wavelets $\mypsi_{\alpha}(t)$ and $\mypsi_{\beta}(\log \lambda)$ of various rates $\alpha$ and scales $\beta$ (resp.). 
    Orientation is determined by the sign of $\beta$, otherwise known as the \emph{spin} variable falling in $\{-1, 1\}$. See Section \ref{sec:jtfs} for details on JTFS.}
    \label{fig:jtfs_filters}
\end{figure}

\subsection{Wavelet scalogram}\label{sec:scalogram}
Let $\boldsymbol{\psi} \in \mathbf{L}^2(\mathbb{R, C})$ be a complex-valued zero-average wavelet filter of unit center frequency and bandwidth $1 / Q_1$. 
We define a constant-$Q$ filterbank of dilations from $\boldsymbol{\psi}$ as $\boldsymbol{\psi}_\lambda: t\longmapsto \lambda\boldsymbol{\psi}(\lambda t)$, with constant quality factor $Q_1$. 
Each wavelet has a centre frequency $\lambda$ and a bandwidth of $\lambda / Q_1$.
We discretise the frequency variable $\lambda$ under a geometric progression of common ratio $2^{\frac{1}{Q_1}}$, starting from $\lambda / Q_1$. 
For a constant quality factor of $Q_1 = 1$, subsequent wavelet centre frequencies are spaced by an octave, i.e. a dyadic wavelet filterbank. 

Convolving the filterbank \mypsi{} with a waveform $\myx \in \mathbf{L}^2(\mathbb{R})$ and applying a pointwise complex modulus gives the wavelet scalogram $\mathbf{U}_1$:
\begin{equation}\label{eqn:u1}
\mathbf{U}_1 \myx(t, \lambda) = |\myx \ast \mypsi_\lambda|(t)
\end{equation}
$\mathbf{U}_1$ is indexed by time and $\log$-frequency, corresponding to the commmonly known constant-Q transform in time--frequency analysis. 

\subsection{Time--frequency wavelets}
Similarly to Section \ref{sec:scalogram}, we define another two wavelets $\mypsi^{\mathrm{t}}$ and $\mypsi^{\mathrm{f}}$ along the time and $\log$-frequency axes, with quality factors equivalent to $Q_2$ and $Q_{\mathrm{fr}}$, respectively. 
We then derive two filterbanks $\mypsi_{\alpha}^\mathrm{t}$ and $\mypsi_{\beta}^\mathrm{f}$, with center frequencies of $\alpha$ and $\beta$, where
\begin{equation}
    \mypsi_{\alpha}^\mathrm{t}(t) = \alpha \mypsi^\mathrm{t} (\alpha t)
\end{equation}
\begin{equation}
\mypsi_{\beta}^\mathrm{f}(\log_2 \lambda) = \beta \mypsi^\mathrm{f} (\beta \log_2 \lambda)
\end{equation}
As in the computation of $\mathbf{U}_1$, we discretize $\alpha$ and $\beta$ by geometric progressions of common ratios $2^\frac{1}{Q_2}$ and $2^\frac{1}{Q_{\mathrm{fr}}}$.
We interpret the frequency variable $\alpha$ and $\beta$ from a perspective of auditory STRFs \cite{chi2005multiresolution}: $\alpha$ is the temporal modulation rate measured in Hz, while $\beta$ is the frequential modulation scale measured in cycles per octave. 

The outer product between $\mypsi^\mathrm{t}_{\alpha}$ and $\mypsi^\mathrm{f}_{\beta}$ forms a family of 2D wavelets of various rates $\alpha$ and scales $\beta$. 
We convolve $\mypsi^\mathrm{t}_{\alpha}$ and $\mypsi^\mathrm{f}_{\beta}$ with $\mathbf{U}_1\bm{x}$ in sequence and apply a pointwise complex modulus, resulting in a four-way tensor indexed ($t$, $\lambda$, $\alpha$, $\beta$):
\begin{equation}\label{eqn:u2}
\mathbf{U}_2 \bm{x}(t, \lambda, \alpha, \beta) = |\mathbf{U}_1 \ast \mypsi_\alpha^\mathrm{t} \ast \mypsi_\beta^\mathrm{f}|
\end{equation}
 
In Fig. \ref{fig:jtfs_filters} we visualize the real part of the 2D wavelet filters in the time--frequency domain. The wavelets are of rate $\alpha$, scale $\beta$ and orientation (upward or downward) along $\log_2 \lambda$, capturing multiscale oscillatory patterns in time and frequency. 

\subsection{Local averaging}
We compute \emph{first-order joint time--frequency scattering} coefficients by convolving the scalogram $\mathbf{U}_1\bm{x}$ of Eqn. (\ref{eqn:u1}) with a Gaussian lowpass filter $\boldsymbol{\phi}_T$ of width $T$, followed by convolution with $\mypsi_\beta$ ($\beta \geq 0$) over the log-frequency axis, then pointwise complex modulus:
\begin{equation}\label{eqn:s1}
\mathbf{S}_1 \bm{x}(t, \lambda, \alpha = 0, \beta) = |\mathbf{U}_1 x(t, \lambda) \ast \boldsymbol{\phi}_T \ast \mypsi_\beta|
\end{equation}
Before convolution with $\mypsi_\beta$, we subsample the output of $\mathbf{U}_1 x(t, \lambda) \ast \boldsymbol{\phi}_T$ along time, resulting in a sampling rate proportional to $1/T$. 
Indeed, Eqn.~(\ref{eqn:s1}) is a special case of Eqn.~(\ref{eqn:u2}) in which modulation rate $\alpha = 0$ by the use of $\boldsymbol{\phi}_T$.

We define the \emph{second-order joint time--frequency scattering} transform of $\bm{x}$ as:
\begin{equation}\label{eqn:s2}
\mathbf{S}_2 \bm{x}(t, \lambda, \alpha, \beta) = \mathbf{U}_2 x(t, \lambda) \ast \boldsymbol{\phi}_T \ast \boldsymbol{\phi}_F
\end{equation}
where $\boldsymbol{\phi}_F$ is a Gaussian lowpass filter over the $\log$-frequency dimension of width $F$. 
For the special case of $\beta = 0$ in Eqn. \ref{eqn:u2}, $\mypsi_\beta$ performs the role of $\boldsymbol{\phi}_F$, yielding: 
\begin{equation}\label{eqn:s2_beta}
\mathbf{S}_2 \bm{x}(t, \lambda, \alpha, \beta = 0) = |\mathbf{U}_1 x(t, \lambda) \ast \mypsi_\alpha^\mathrm{t} \ast \boldsymbol{\phi}_F| \ast \boldsymbol{\phi}_T
\end{equation}
In both Eqns. (\ref{eqn:s2}) and (\ref{eqn:s2_beta}), we subsample $\mathbf{S}_2\bm{x}$ to sampling rates of $T^{-1}$ and $F^{-1}$ over the time and $\log$-frequency axes, respectively.  
Lowpass filtering with $\boldsymbol{\phi}_T$ and $\boldsymbol{\phi}_F$ provides invariance to time shifts and frequency transpositions up to a scale of $T$ and $F$ respectively.
The combination of $\mathbf{S}_1\bm{x}$ and $\mathbf{S}_2\bm{x}$, i.e. $\mathbf{S}\bm{x}=\{\mathbf{S}_1\bm{x},\mathbf{S}_2\bm{x}\}$, allows us to cover all paths combining the variables $(\lambda, \alpha, \beta)$.
In Section \ref{sec:dmo} we introduce the use of $\mathbf{S}\bm{x}$ as a DTFA operator for mesostructures.

In Fig. \ref{fig:chirps}, we highlighted the need for a operator that models mesostructures. The stream of chirplets is displaced in frequency at a particular rate. At second-order, JTFS describes the larger scale spectrotemporal structure that is not captured by $\mathbf{S}_1$. Moreover, JTFS is time-invariant, making it a reliable measure of mesostructural similarity up to time scale $T$.
\begin{figure}
    \centering
    \begin{tabular}{cc}
      JTFS & MSS \\
      \includegraphics[width=
      0.45\columnwidth]{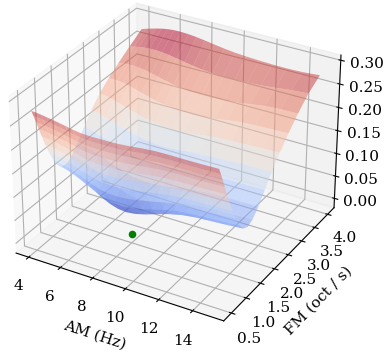} & 
      \includegraphics[width=0.45\columnwidth]{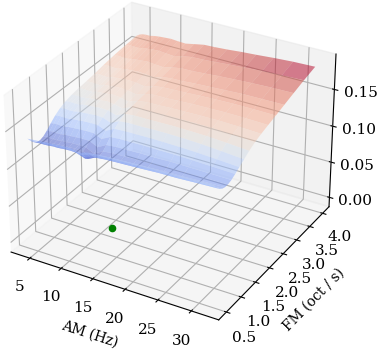} \\
      \includegraphics[width=0.45\columnwidth]{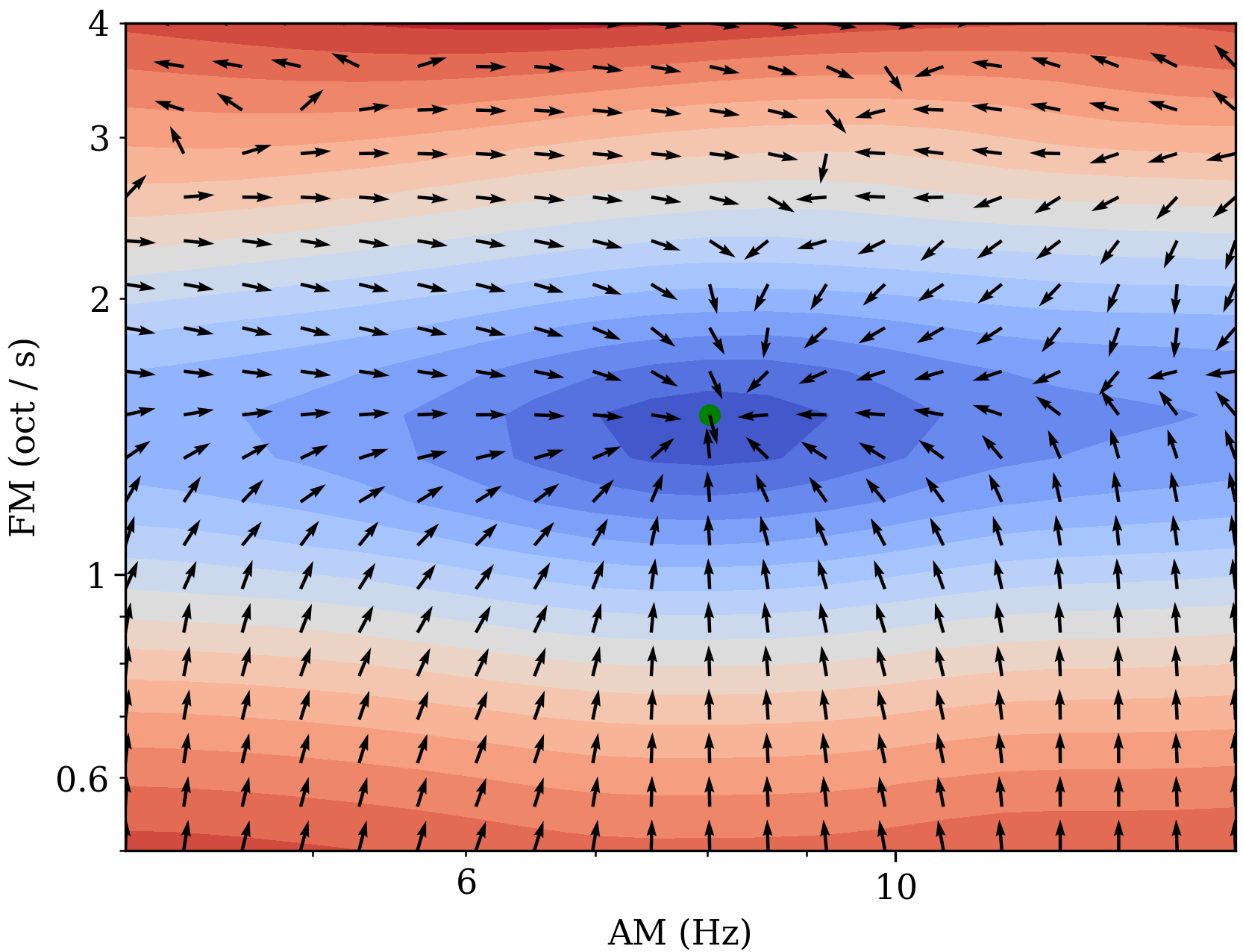} &
      \includegraphics[width=0.45\columnwidth]{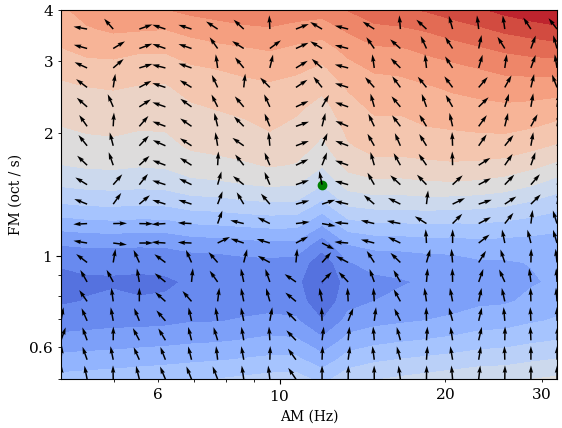}
    \end{tabular}
    \caption{Loss surface and gradient field visualization under $\myPhi$ as JTFS (left) and MSS (right) for sounds synthesized by \myg{} (see Section \ref{sec:am_chirp}). Sounds are sampled from a logarithmically spaced grid on $f_\mathrm{m}$ and $\gamma$. 
    We plot the target sound as a green dot and compute the loss between the target and a sound generated at every point on the grid. 
    We time shift the generated sound relative to the target by a constant of $\tau=2^{10}$ samples. 
    In the quiver plots, we evaluate the gradient of the loss operator with respect to synthesis parameters $f_{\mathrm{m}}$ and $\gamma$. 
    The direction of the arrows is indicative of the informativeness of the distance computed on $\myPhi \circ \myg$ with respect to \mytheta{}. 
    In the case of $\myPhi_{\mathrm{JTFS}}$, we observe a 3D loss surface whose global minimum is centred around the target sound, while gradients point towards the target. Contrarily, the global minimum of $\myPhi_{\mathrm{MSS}}$ does not centre around the target or reach 0. 
    In the presence of small time shifts, the MSS loss appears insensitive to differences in AM and uninformative with respect to $\mytheta$.}
    \label{fig:gradient_plot}
\end{figure}

\section{DIFFERENTIABLE MESOSTRUCTURAL OPERATOR} \label{sec:dmo}
\begin{figure}
    \centering
    \includegraphics[width=0.9\columnwidth]{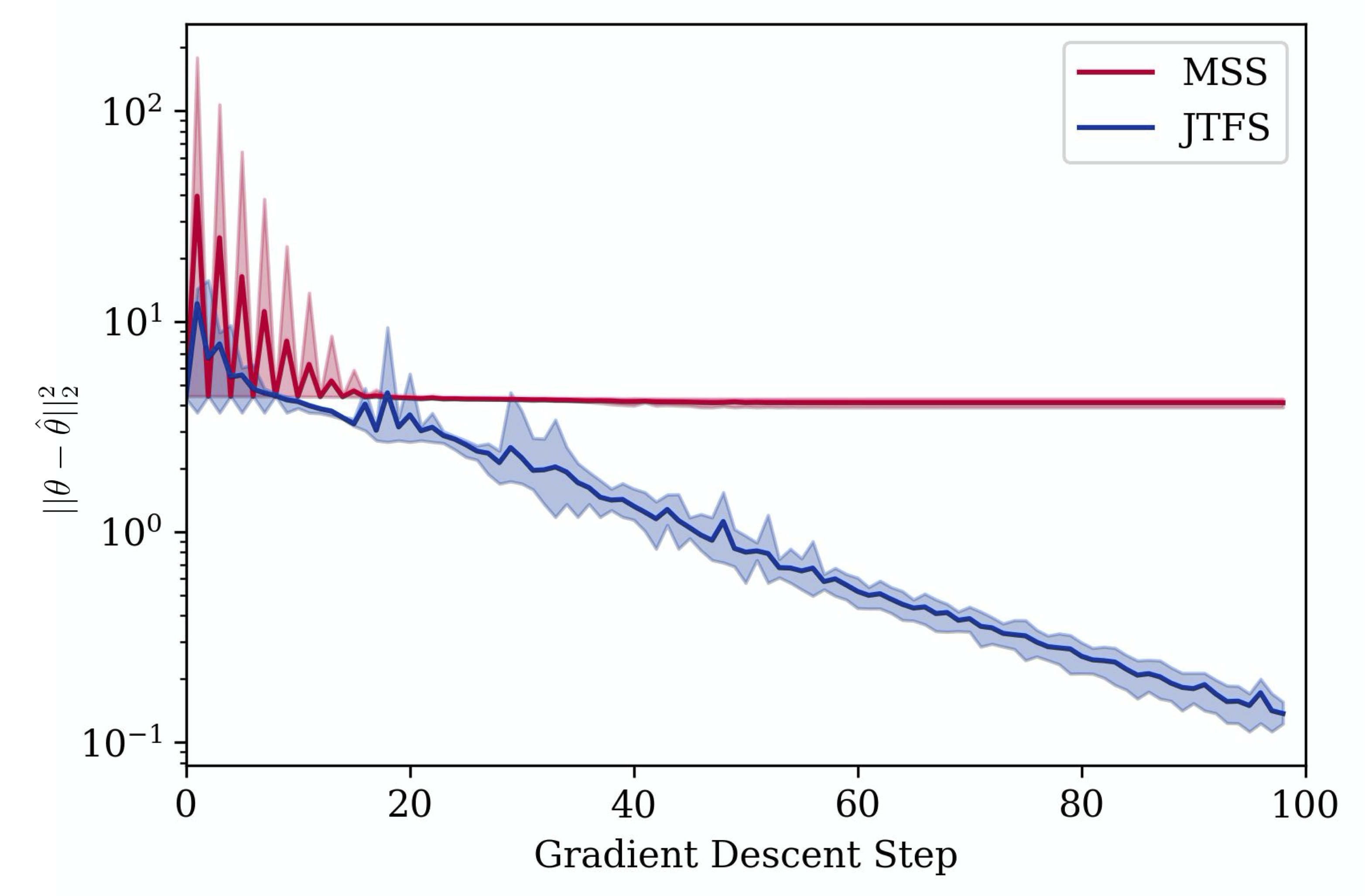}
    \caption{Parameter distance $||\mytheta - \mythetatilde||_2$ over gradient descent iterations with \myPhi{} as MSS and JTFS.
    The target sound has parameters $\mytheta = [8.49, 1.49]$. We initialize the predicted sound at $\mythetatilde_0 = [4, 0.5]$. 
    The line plots the mean distance at each iteration for multiple runs that shift the predicted sample in time by $\tau = \{2^2, 2^4, 2^7, 2^{10}\}$ samples.
    The shaded region indicates the range across different time shifts.}
    \label{fig:gradient_plot_iters}
\end{figure}

\begin{figure}
    \centering
      \includegraphics[width=\columnwidth]{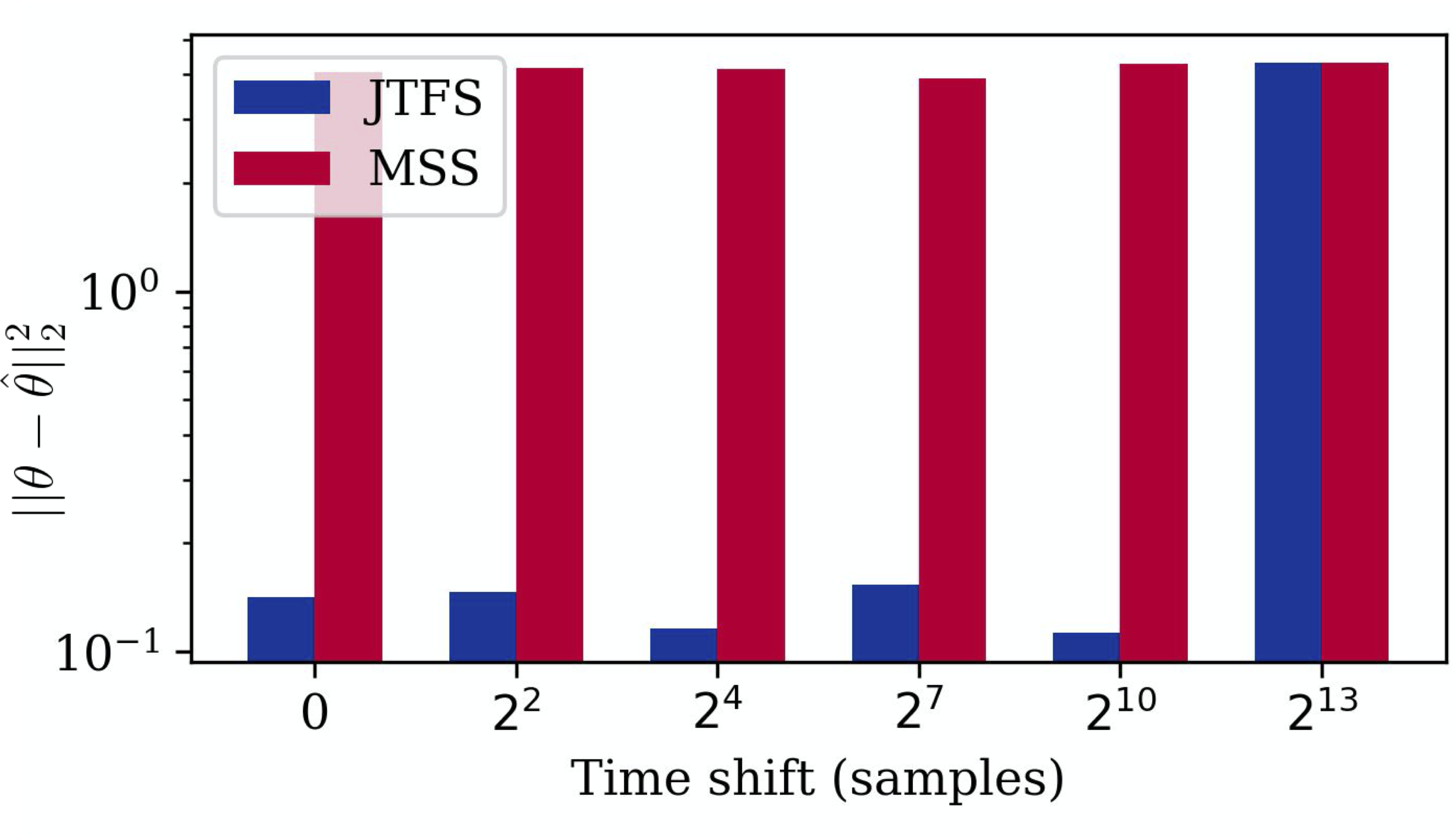} \\
    \caption{Final parameter distance $||\mytheta - \mythetatilde||_2$ after gradient descent for $\myg(\mytheta)(t)$ and $\myg(\mythetatilde)(t - \tau)$, for $\mytheta = [8.49, 1.49]$, $\mythetatilde_0 = [4, 0.5]$. Each run (x-axis) is optimized under a different time shift $\tau$ on the predicted audio. 
    JTFS is invariant up to the support $T = 2^{13}$ of its lowpass filter. 
    We observe that convergence in parameter recovery is stable to time shifts under our differentiable mesostructural operator $\myPhi \circ \myg$, in the case that $\myPhi$ is JTFS. 
    Optimization is unstable when $\myPhi$ is a spectrogram operator.}
    \label{fig:ploss_time_shifts}
\end{figure}

\begin{figure*}
    \centering
    \begin{tabular}{c}
    start far from target \hspace{3cm}
    start near target \hspace{3cm}
    start anywhere
    \\
    \includegraphics[width=\textwidth]{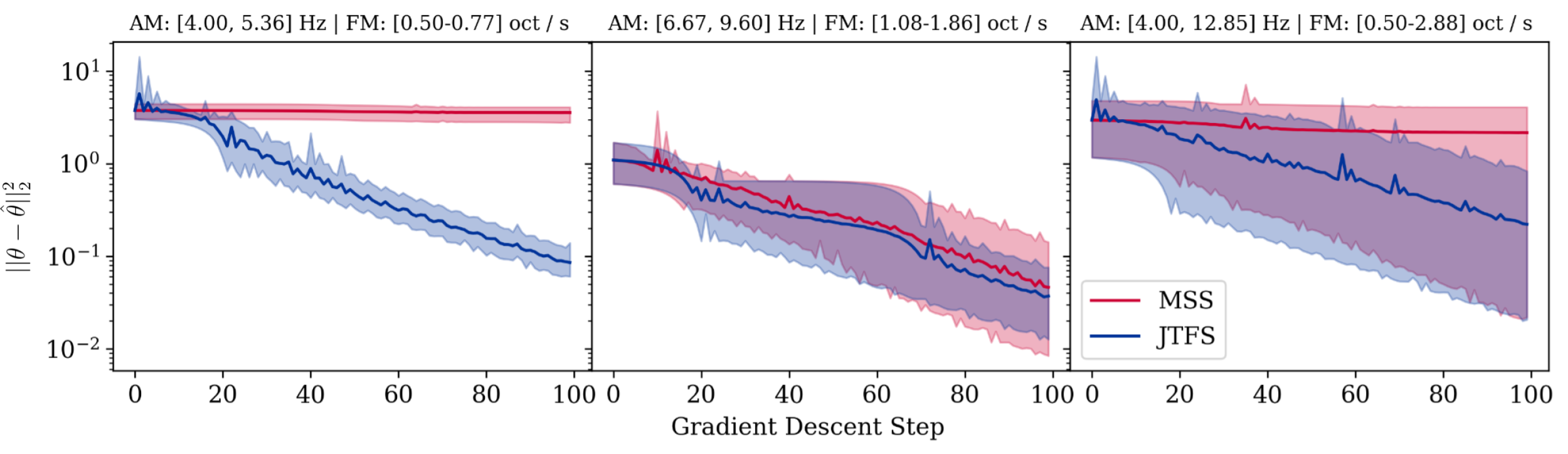}
    \end{tabular}
    \caption{Parameter distance $||\mytheta - \mythetatilde||$ (log-scale) over gradient descent iterations with \myPhi{} as MSS and JTFS in 3 scenarios: (left) 5 initialisations of $\mythetatilde$ far from the target, (centre) 5 initialisations of $\mythetatilde$ in the neighbourhood of the target and (right) 5 initialisations of $\mythetatilde$ across the range of the grid. We do not apply a time shift to the predicted sound (see Fig. \ref{fig:gradient_plot} for gradient visualisation).
    The target sound has parameters $\mytheta = [8.49, 1.49]$.
    The lines indicate the mean distance at each iteration across 5 runs of different $\mythetatilde$ initialisation.
    The shaded region indicates the range across the 5 initialisations. The titles indicate the range of the initial $\mythetatilde$. We highlight that even with no time shifts, MSS only recovers $\mytheta$ well when $\mythetatilde$ is initialised in its local neighbourhood (centre). When $\mythetatilde$ is initialised far from the target (left), MSS fails to converge. Starting anywhere (right), converges in the best case, but on average fails to converge and is close to the worst case.}
    \label{fig:grad_plot_inits}
\end{figure*}
\begin{figure*}
    \centering
    \begin{tabular}{cccc}
    JTFS & MSS & JTFS & MSS \\
    no time shift & no time shift & random time shift & random time shift \\
    \includegraphics[width=0.45\columnwidth]{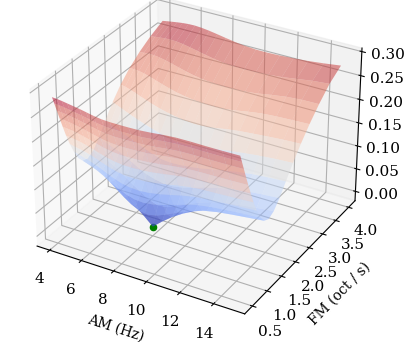} 
    & \includegraphics[width=0.45\columnwidth]{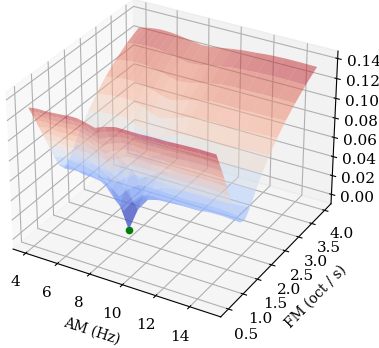} 
    & \includegraphics[width=0.45\columnwidth]{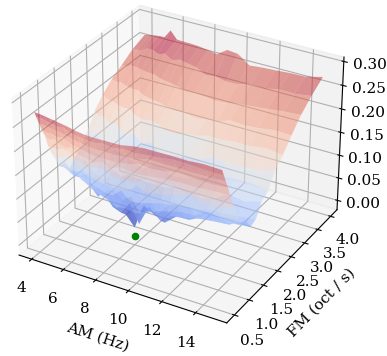}
    & \includegraphics[width=0.45\columnwidth]{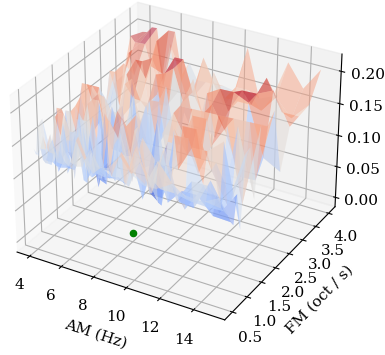} \\ \\\newline
   \includegraphics[width=0.45\columnwidth]{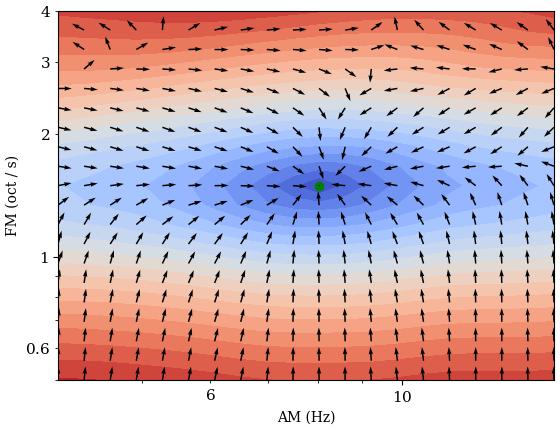}
    & \includegraphics[width=0.45\columnwidth]{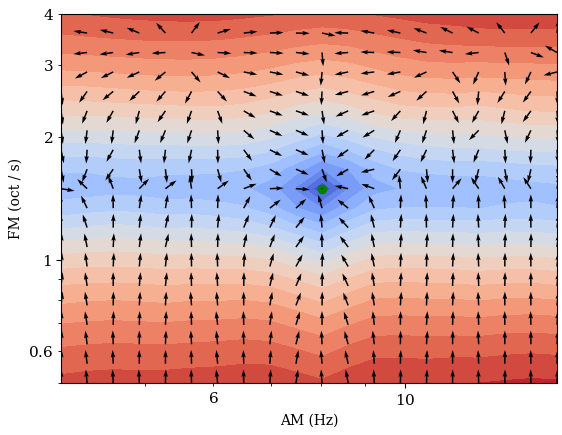}
    & \includegraphics[width=0.45\columnwidth]{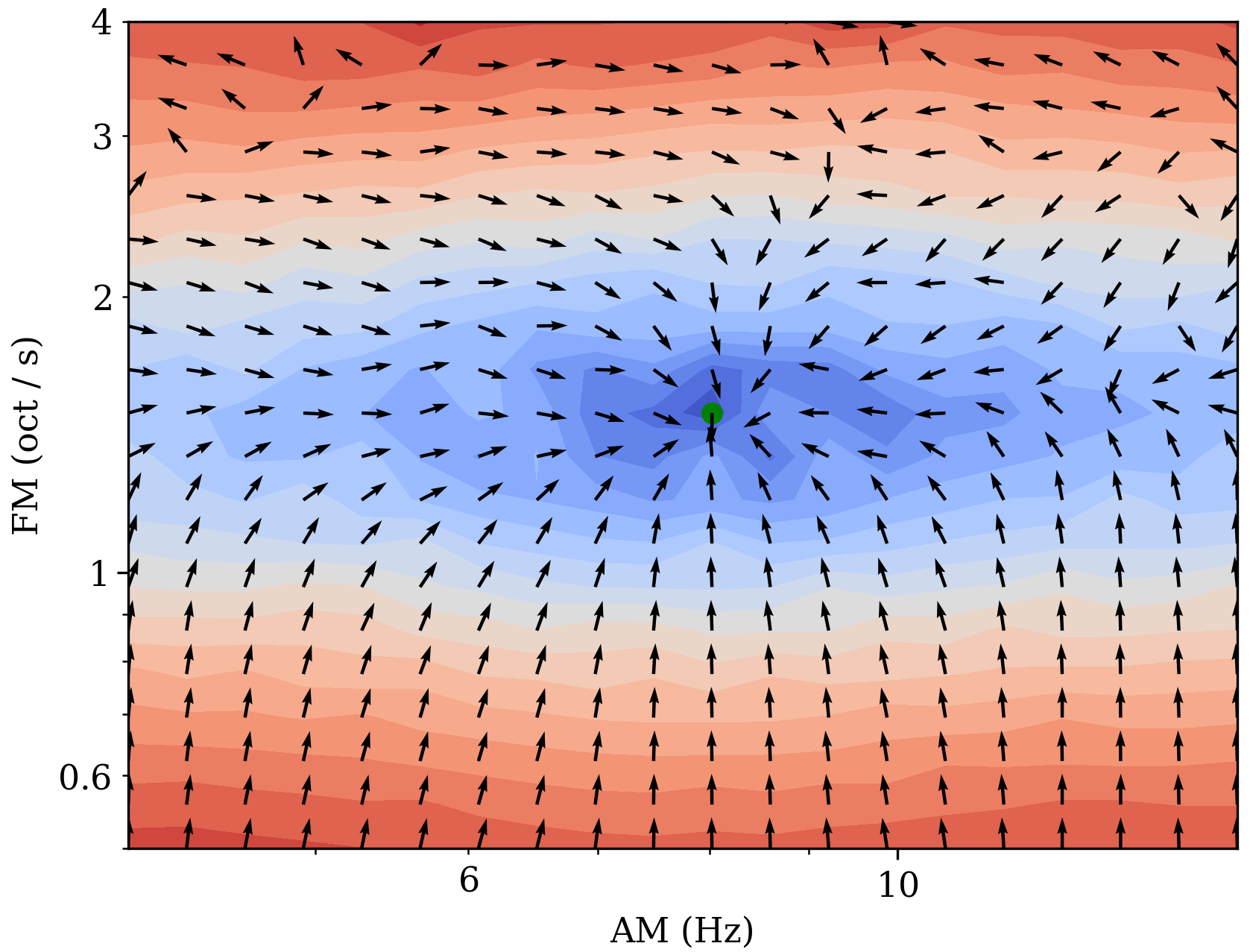}
    &   \includegraphics[width=0.45\columnwidth]{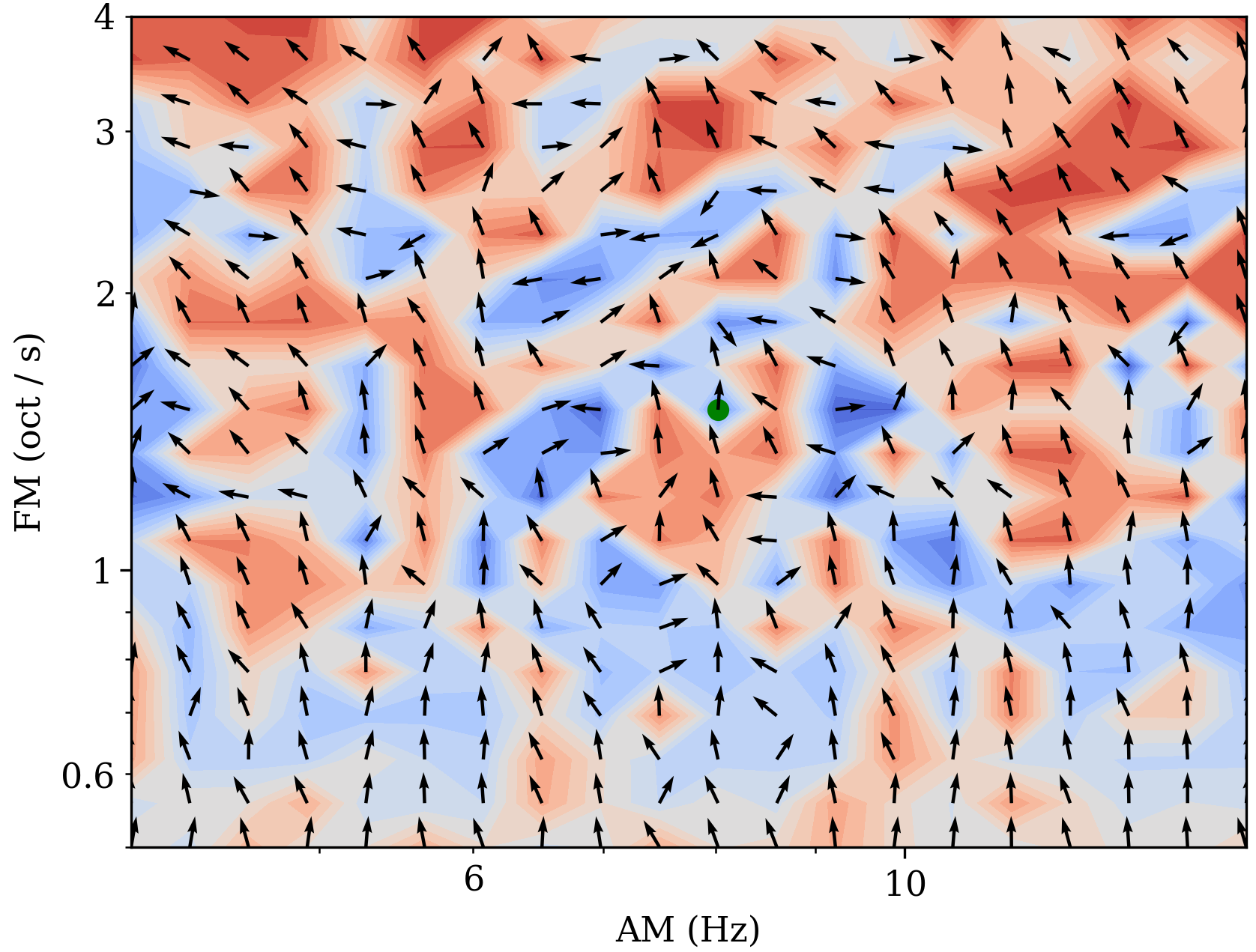}
    \end{tabular}
    \caption{Loss surfaces (top) and gradient fields (bottom) under $\myPhi_{\mathrm{JTFS}}$ and the $\myPhi_{\mathrm{MSS}}$ for sounds synthesized by \myg{} (see Section \ref{sec:am_chirp}), sampled from a logarithmically spaced grid on $f_\mathrm{m}$ and $\gamma$. Each sound is randomly shifted in time relative to the target by $2^n$ samples, where $n$ is sampled uniformly between $[8, 12]$.
    We plot the target sound as a green dot and compute the loss under $\myPhi_{\mathrm{JTFS}}$ and $\myPhi_{\mathrm{MSS}}$ between each sound and the target. 
    In the quiver plots, we evaluate the gradient of the loss operator with respect to the synthesis parameters $f_{\mathrm{m}}$ and $\gamma$ of the generated sound. 
    In the case of both no time shifts, JTFS gradients point towards the target and the distance around 0 when is at the target. Without time shifts, MSS computes distance between objects that intersect in the time--frequency domain. Its gradients appear to lead to the target, however it suffers from local minima along AM, as demonstrated by convergence in Fig.~ \ref{fig:grad_plot_inits}. In the presence of random time shifts, JTFS is appears robust while MSS is highly unstable and prone to local minima.}
    \label{fig:gradient_plot_time_shifts}
\end{figure*}
In this section, we introduce a \emph{differentiable mesostructural operator} for time--frequency analysis. 
Such an operator is needed in optimization scenarios that require a differentiable measure of similarity, such as autoencoding. 

In Section \ref{sec:am_chirp}, we defined a differentiable arpeggiator \myg{} whose parameters \mytheta{} govern the \emph{mesostructure} in \myx{}. 
We now seek a differentiable operator $\myPhi \circ \myg$ that provides a model to control the low-dimensional parameter space \mytheta{}.  
By way of distance and gradient visualization under $\myPhi \circ \myg$, we set out to assess the suitability of \myPhi{} for modelling \mytheta{} in a sound matching task. 

We consider two DTFA operators in the role of \myPhi{}: (i) the multiscale spectrogram (MSS) (approximately $\mathbf{U}_1\bm{x}$) and (ii) time--frequency scattering ($\mathbf{S}\bm{x}=\{\mathbf{S}_1\bm{x},\mathbf{S}_2\bm{x}\}$) (JTFS). 
In case (i), we deem a small distance between two sounds to be an indication of same \emph{microstructure}. On the contrary, similarity in case (ii) suggests the same \emph{mesostructure}. 
Although identical $\mathbf{U}_1$ implies equality in mesostructure, the reverse is not true, i.e. in the case of time shifts and non-stationary frequency. 

Previously, JTFS has offered assessment of similarity between musical instrument playing techniques that underlie mesostructure.
With the DTFA operator \myPhi{}, there is potential to model mesostructures by their similarity as expressed in terms of the raw audio waveform, synthesis parameters or neural network weights. 
In cases such as granular synthesis, it may be desirable to control mesostructure, while allowing microstructure to stochastically vary.

\subsection{Gradient computation \& visualization} \label{sec:grad-viz}
We evaluate a distance objective under the operator $\myPhi \circ \myg$ as a proxy for distance in \mytheta:
\begin{equation} \label{eqn:loss}
    \mathcal{L}_{\mytheta} (\tilde{\mytheta})=\Vert 
    (\myPhi \circ \myg)(\mytheta) - (\myPhi \circ \myg)(\tilde{\mytheta})
    \Vert_2^2
\end{equation}
For a given parameter estimate $\tilde{\mytheta}$, the gradient $\nabla\mathcal{L}_{\theta}$ of the distance to the target \mytheta{} is:
\begin{align} \label{grad_decomp}
\nabla \mathcal{L}_{\theta}(\tilde{\theta})
    & = -2\Big((\myPhi \circ \myg)(\mytheta) - (\myPhi \circ \myg)(\tilde{\mytheta})\Big)^{T} \cdot \nabla (\myPhi \circ \myg) (\tilde{\theta})
\end{align}
The first term in Eqn. (\ref{grad_decomp}) is a row vector of length $P=\dim\Big((\myPhi \circ \myg)(\mytheta)\Big)$ and the second term is a matrix of dimension $P \times \dim(\tilde{\mytheta})$. The dot product between the row vector in the first term and each column vector in the high-dimensional Jacobian matrix $\nabla (\myPhi{} \circ \myg)$ yields a low-dimensional vector of $\dim(\mytheta{})$. Each column of the Jacobian matrix can be seen as the direction of steepest descent in the parameter space, such that distance in $\myPhi$ is minimized. Therefore the operator $\myPhi \circ \myg$ should result in distances that reflect sensitivity and direction of changes in \mytheta{}.


In $\mathcal{L}_\theta$ of Eqn. (\ref{eqn:loss}), we adopt time--frequency scattering ($\boldsymbol{S} \myx$) (see Section~\ref{sec:jtfs}) in the role of $\myPhi$. Otherwise, we refer to $\mathcal{L}_{\mytheta}^{MSS}$ when using the multi-scale spectrogram (MSS). In the JTFS transform, we set $J = 12$, $J_{fr} = 5$, $Q_1 = 8$, $Q_2 = 2$, $Q_{fr} = 2$, and set $F = 0$ to disable frequency averaging. 
 
 Alternatively, we refer to $\mathcal{L}_{\mytheta}^{MSS}$ when using the multi-scale spectrogram (MSS). Let $\myPhi_{\mathrm{STFT}}^{(n)}$ be the short-time fourier transform coefficients computed with a window size of $2^n$. 
We compute the MSS loss in Eqn. (\ref{eqn:mse-msstft}), which is the average of L1 distances between spectrograms at multiple STFT resolutions:
\begin{equation}\label{eqn:mse-msstft}
    \mathcal{L}_{\mytheta}^{MSS}(\tilde{\mytheta}) = \frac{1}{N}\sum_{i=5}^{10} 
    \vert (\myPhi_{\mathrm{STFT}}^{(n)} \circ \myg)(\mytheta) - (\myPhi_{\mathrm{STFT}}^{(n)} \circ \myg)(\tilde{\mytheta})\vert
\end{equation}
The chosen resolutions account for the sampling rate of 8192 Hz used by \myg{}.
We set $w = 2$ octaves in all subsequent experiments and normalize the amplitude of each $\boldsymbol{g_\theta}$.

For this experiment, we uniformly sample a grid of $20 \times 20$ AM/FM rates $(\boldsymbol{f_\mathrm{m}}, \boldsymbol{\gamma})$ on a log-scale ranging from 4 to 16 Hz and 0.5 to 4 octaves per second, leading to 400 signals with a carrier frequency of $\boldsymbol{f_\mathrm{c}} = 512 \;\text{Hz}$. We designate the centre of the grid $\boldsymbol{f_\mathrm{m}} = 8.29 \; \text{Hz}$ and $\boldsymbol{\gamma} = 1.49\; 
 \text{octaves / second}$ as the target sound.
We introduce a constant time shift $\tau = 2^{10}$ samples to the target sound in order to test the stability of gradients under perturbations in microstructures.
We evaluate $\mathcal{L}_{\mytheta}$ and $\nabla \mathcal{L}_{\mytheta}$ associated to each sound for the two DTFA operators $\myPhi_{\mathrm{STFT}}$ and $\myPhi_{\mathrm{JTFS}}$. 

We visualize the loss surfaces and gradient fields with respect to $\mythetatilde$ in Fig. \ref{fig:gradient_plot}. We observe that the JTFS operator forms a loss surface with a single local minimum that is located at the target sound's \mytheta{}. Meanwhile gradients across the sampled parameters $\mythetatilde$ consistently point towards the target, despite certain exceptions at high $\gamma$, which acoustically correspond to very high FM rate.
Contrarily, MSS loss gradient suffers from multiple local minima and does not reach the global minimum when $\mythetatilde$ is located at the target due to time shift equivariance. We highlight that the MSS distance is insensitive to variation along AM, making it unsuitable for modelling mesostructures.

In line with our findings, previous work \cite{muradeli2022differentiable} found that 3D visualizations of the manifold embedding of JTFS' nearest neighbour graph revealed a 3D mesh whose principal components correlated with parameters describing
carrier frequency, AM and FM. 
Moreover, $K$-nearest neighbours regression using a nearest neighbours graph in JTFS space produced error ratios close to unity for each of the three parameters. 

\subsection{Sound Matching by gradient descent}
Unlike classic sound matching literature, where $\tilde{\mytheta{}}$ is estimated from a forward pass through trainable $\myfW$ (i.e., neural network weights), we formulate sound matching as an inverse problem in $(\myPhi{} \circ \myg{})$. For the sake of simplicity, we do not learn any weights to approximate $\mytheta{}$. 

Using the gradients derived in Section \ref{sec:grad-viz}, we attempt sound matching of a target state in \mytheta{} using a simple gradient descent scheme with bold driver heuristics.
We perform additive updates to $\tilde{\mytheta}$ along the direction dictated by gradient $\nabla_{\tilde{\mytheta}}\mathcal{L}_{\mytheta}$: 
\begin{equation}
\tilde{\mytheta} \leftarrow \tilde{\mytheta} - \alpha \nabla_{\tilde{\mytheta}} \mathcal{L}_{\mytheta}
\end{equation}
Our bold driver heuristic increases the learning rate $\alpha$ by a factor of 1.2 when $\mathcal{L}_{\mytheta}$ decreases it by a factor of 2 otherwise. Our evaluation metric in parameter space is defined as:
\begin{equation}
    \mathcal{L}_{\mytheta}( \mythetatilde) = \Vert \mytheta - \mythetatilde\Vert^2_2
\end{equation}

Fig. \ref{fig:gradient_plot_iters} shows the mean L2 parameter error over gradient descent steps for each $\myPhi{}$. We select a fixed target and initial prediction. We run multiple optimizations that consider time shifts between $0$ and $2^{10}$ samples on the target audio.
Across time-shifts within the support $T$ of the lowpass filter in $\myPhi{}_{JTFS}$, convergence is stable and reaches close to 0. We observe that MSS does not converge and $\mathcal{L}_{\mytheta}( \mythetatilde)$ does not advance far from its initial value, including the case of no time shifts. In Fig. \ref{fig:ploss_time_shifts}, we further illustrate the effects of time shifts for DTFA, validating that JTFS is a time-invariant mesostructural operator up to support $T$.

\subsection{Time invariance}
In Fig. \ref{fig:grad_plot_inits}, we explore the gradient convergence for different initialisations of $\mythetatilde$ but \emph{without} time shifting the predicted sound. 
In each plot, we perform gradient descent for 5 different initialisations of $\mythetatilde$: (i) far away from the target sound, (ii) in the local neighbourhood of the target sound and (iii) broadly across the parameter grid. We highlight that JTFS is able to converge to the solution in each of the 3 initialisation schemes, as corroborated by its gradients in Fig. \ref{fig:gradient_plot_time_shifts}. We observe that even without time shifts, MSS fails to recover the target sound in the case that the parameter initialisation is far from the target. 
MSS does indeed recover the target sound if $\mythetatilde$ is initialised in the neighbourhood of the target. Although when starting anywhere, MSS does indeed converge in the best case, but on average it is close to the worst case which does not converge.

Fig. \ref{fig:gradient_plot_time_shifts} shows the loss surface and gradient fields for $\myPhi_{\mathrm{JTFS}}$ and $\myPhi_{\mathrm{MSS}}$ with no time shifts and random time shifts applied to the predicted sound.
Despite MSS reaching the global minimum when the predicted sound is centred at the target, our experiments in gradient descent demonstrate that it is only stable when $\mythetatilde$ is initialised within the local region of the target $\boldsymbol{\theta}$. 
When we apply a random time shift to the predicted sound, the MSS loss is highly unstable and produces many local minima that are not located at the target sound. 
As expected, the JTFS gradient is highly stable with no time shifts. Even in the presence of random time shifts, JTFS is an invariant representation of spectrotemporal modulations upto time shifts $T$.

\section{CONCLUSION}
Differentiable time--frequency analysis (DTFA) is an emerging direction for audio deep learning tasks. 
The current state-of-the-art for autoencoding, audio restoration and sound matching predominantly perform DTFA in the spectrogram domain. 
However, spectrogram loss suffers from numerical instabilities when computing similarity in the context of: (i) time shifts beyond the scale of the spectrogram window and (ii) nonstationarity that arises from synthesis parameters. 
These prohibit the reliability of spectrogram loss as a similarity metric for modelling multiscale musical structures.

In this paper, we introduced the \emph{differentiable mesostructural operator}, comprising of DTFA and an arpeggio synthesiser, for time--frequency analysis of mesostructures.
We model synthesis parameters for a sound matching task using the joint time--frequency scattering (JTFS) for DTFA of structures that are identifiable beyond the locality of microstructure; i.e. amplitude and frequency modulations of a chirplet synthesizer.
Notably, JTFS offers a differentiable and scalable implementation of: auditory spectrotemporal receptive fields, multiscale analysis in the time--frequency domain and invariance to time shifts. 

However, despite prior evidence that JTFS accurately models similarities in signals containing spectrotemporal modulations, JTFS is yet to be assessed in DTFA for inverse problems and control in sound synthesis.
By analysis of the gradient of our DTFA operator with respect to synthesis parameters, we showed that in contrast to spectrogram losses, JTFS distance is suitable for modelling similarity in synthesis parameters that describe mesostructure. 
We demonstrated the stability of JTFS as a DTFA operator in sound matching by gradient descent, particularly in the case of time shifts.

This work lays the foundations for further experiments in DTFA for autoencoding, sound matching, resynthesis and computer music composition. 
Indeed, our differentiable mesostructural operator could be used as a model of the raw audio waveform directly, however this approach is prone to resynthesis artifacts \cite{engel2017neural, lostanlenflorian19}. 
We have shown that by means of DTFA, we can model low-dimensional synthesis parameters that shape sequential audio events. 
A direction for future work lies in differentiable parametric texture synthesis, in which texture similarity may be optimized in terms of parameters that derive larger scale structures; e.g. beyond the definition of individual grains in granular synthesis.  
\section{ACKNOWLEDGMENT}
Cyrus Vahidi is a researcher at the UKRI CDT in AI and Music, supported jointly by the UKRI (grant number EP/S022694/1) and Music Tribe. This work was conducted during a research visit at LS2N, CNRS.
Changhong Wang is supported by an Atlanstic2020 project on Trainable Acoustic Sensors (TrAcS).
\Urlmuskip=0mu plus 1mu\relax
\bibliography{main.bbl}
\bibliographystyle{jaes.bst}





\end{document}